\documentclass[runningheads]{llncs}
\usepackage[T1]{fontenc}
\usepackage{microtype}
\usepackage{graphicx}
\usepackage{hyperref}
\usepackage{color}

\usepackage{latexsym,amsmath,amssymb}
\usepackage{mathpartir}
\usepackage{stackrel}
\usepackage[only,owedge,llfloor,rrfloor,lightning,lbag,rbag,Longmapsto,llbracket,rrbracket]{stmaryrd}
\usepackage{wrapfig,subcaption}
\usepackage{relsize}
\usepackage{thmtools} % for restatable theorems, to put proofs in appendix.
\usepackage{thm-restate}
\RequirePackage{listings}
\definecolor{dkgreen}{rgb}{0,0.6,0}
\definecolor{gray}{rgb}{0.5,0.5,0.5}
\definecolor{mauve}{rgb}{0.58,0,0.82}
\definecolor{navy}{rgb}{0,0,0.5}

\lstdefinelanguage{whyrel}
{
basicstyle=\linespread{0.5}\footnotesize\ttfamily,%
keywordstyle=[1]\color{blue},%
morekeywords=[1]{axiom,biproc,interface,bimodule,new,%
class,do,%
else,end,exists,fi,function,%
if,import,in,lemma,left,%
module,not,od,old,in,relator,%
boundary,meth,type,%
proc,predicate,private,privateInv,public,result,right,bipred,biproc,biassert,biassume%
returns,then,var,while,whyimport,let,done%
},%
keywordstyle=[2]\color{mauve},%
morekeywords=[2]{rgn,any,ghost,forall,Both,empty,alloc,mutable},
keywordstyle=[3]\color{navy},%
morekeywords=[1]{assert, assume, requires, ensures, writes, reads, rd, rw, effects, coupling, invariant,inv,variant, connect, with},
string=[b]",%
morekeywords=[2]{true,false},%
commentstyle=\itshape\color{dkgreen},%
columns=[l]fullflexible,%
sensitive=true,%
morecomment=[s]{/*}{*/},%
moredelim=[is][\it\color{dkgreen}]{(*}{*)},
numberstyle=\color{gray},
stringstyle=\color{mauve},
emphstyle=[1]{\color{red}},
emph=[1]{Array, Algebriac, true, false},
escapeinside={*?}{?*},%
keepspaces=true,%
literate=%
 {*<|}{$\lexop$}{1} % note: right does not have *; '*' avoids conflict with '<'
 {*<]}{$\lexcl$}{1}
 {[>}{$\rexop$}{1}
 {|>}{$\rexcl$}{1}
 {=:=}{$\eqbi$}{1}%
 {'a}{$\alpha$}{1}%
 {'b}{$\beta$}{1}%
 {<}{$<$}{1}%
 {>}{$>$}{1}%
 {<=}{$\le$}{1}%
 {>=}{$\ge$}{1}%
 {<>}{$\ne$}{1}%
 {/\\}{$\land$}{1}%
 {\\/}{ $\lor$ }{3}%
 {\ or(}{ $\lor$(}{3}%
 {not\ }{$\lnot$ }{2}%
 {not(}{$\lnot$(}{2}%
 {+->}{\texttt{+->}}{3}%
 {+->}{$\mapsto$}{2}%
 {-->}{$\Longrightarrow$}{2}%
 {->}{$\Rightarrow$}{2}%
 {<-}{$\leftarrow$}{2}%
 {<->}{$\Leftrightarrow$}{2}%
 {<<}{$\subseteq$}{2}
 {iin}{$\in$}{2}
 {**}{$\union$}{2}
 {\^\^}{$\cap$}{2}
 {\`}{$\Img$}{2}
 {|_}{\mbox{$\lfloor$}}{2}
 {_|}{\mbox{$\rfloor$}}{2}
 {forall}{$\forall$}{2}
 {exists}{$\exists$}{2}
 {agree}{$\Agr$}{2}
 {both}{$\mathbb{B}$}{2}
 {empty}{$\emptyset$}{2}
 {msigma}{$\sigma$}{1}
 {mtau}{$\tau$}{1}
 {mpi}{$\pi$}{1}
 {mrho}{$\rho$}{1}
 {aP}{$\mathcal{P}$}{1}
 {aPP}{$\mathcal{P}'$}{1}
}

\lstdefinelanguage{why3}
{
basicstyle=\linespread{0.5}\footnotesize\ttfamily,%
keywordstyle=[1]\color{blue},%
morekeywords=[1]{axiom,biproc,interface,bimodule,new,%
class,do,%
else,end,exists,fi,function,%
if,import,in,lemma,left,%
module,not,od,old,in,relator,%
boundary,meth,type,%
proc,predicate,private,privateInv,public,result,right,bipred,biproc,biassert,biassume%
returns,then,var,while,whyimport,let,done%
},%
keywordstyle=[2]\color{mauve},%
morekeywords=[2]{rgn,any,ghost,forall,Both,empty,alloc,mutable},
keywordstyle=[3]\color{navy},%
morekeywords=[1]{assert, assume, requires, diverges, ensures, writes, reads, effects, coupling, invariant,inv,variant, connect, with},
string=[b]",%
morekeywords=[2]{true,false},%
commentstyle=\itshape\color{dkgreen},%
columns=[l]fullflexible,%
sensitive=true,%
morecomment=[s]{(*}{*)},%
numberstyle=\color{gray},
stringstyle=\color{mauve},
emphstyle=[1]{\color{red}},
emph=[1]{Array, Algebriac, true, false},
escapeinside={*?}{?*},%
keepspaces=true,%
literate=%
 {*<|}{$\lexop$}{1} % note: right does not have *; '*' avoids conflict with '<'
 {*<]}{$\lexcl$}{1}
 {[>}{$\rexop$}{1}
 {|>}{$\rexcl$}{1}
 {=:=}{$\eqbi$}{1}%
 {'a}{$\alpha$}{1}%
 {'b}{$\beta$}{1}%
 {<}{$<$}{1}%
 {>}{$>$}{1}%
 {<=}{$\le$}{1}%
 {>=}{$\ge$}{1}%
 {<>}{$\ne$}{1}%
 {/\\}{$\land$}{1}%
 {\\/}{ $\lor$ }{3}%
 {\ or(}{ $\lor$(}{3}%
 {not\ }{$\lnot$ }{2}%
 {not(}{$\lnot$(}{2}%
 {+->}{\texttt{+->}}{3}%
 {+->}{$\mapsto$}{2}%
 {-->}{$\Longrightarrow$}{2}%
 {->}{$\rightarrow$}{2}%
 {<-}{$\leftarrow$}{2}%
 {<->}{$\Leftrightarrow$}{2}%
 {<<}{$\subseteq$}{2}
 {iin}{$\in$}{2}
 {**}{$\union$}{2}
 {\^\^}{$\cap$}{2}
 {\`}{$\Img$}{2}
 {|_}{\mbox{$\lfloor$}}{2}
 {_|}{\mbox{$\rfloor$}}{2}
 {forall}{$\forall$}{2}
 {exists}{$\exists$}{2}
 {agree}{$\Agr$}{2}
 {both}{$\mathbb{B}$}{2}
 {empty}{$\emptyset$}{2}
 {msigma}{$\sigma$}{1}
 {mtau}{$\tau$}{1}
 {mpi}{$\pi$}{1}
 {mrho}{$\rho$}{1}
 {aP}{$\mathcal{P}$}{1}
 {aPP}{$\mathcal{P}'$}{1}
}

\lstnewenvironment{whyrel}{\lstset{language=whyrel}}{}

\newcommand{\code}[1]{\mbox{\lstinline[language=whyrel,basicstyle=\ttfamily\footnotesize]{#1}}}

\lstdefinelanguage{whythree}
{
basicstyle=\linespread{0.5}\normalsize\ttfamily,%
keywordstyle=[1]\color{blue},%
morekeywords=[1]{axiom,biproc,bimodule,new,%
class,do,%
else,end,exists,fi,function,%
if,import,in,lemma,left,%
module,not,od,old,in,relator,%
done,begin,%
boundary,meth,%
proc,predicate,private,privateInv,public,result,right,bipred,biproc,biassert,biassume%
returns,then,var,while,whyimport,let,any%
},%
keywordstyle=[2]\color{mauve},%
morekeywords=[2]{rgn,ghost,Algebraic,forall,Both,empty,alloc,state,refperm},
keywordstyle=[3]\color{navy},%
morekeywords=[1]{assert, assume, requires, ensures, writes, returns,reads,invariant,variant, connect, with},
morekeywords=[3]{ref},
string=[b]",%
morekeywords=[2]{true,false},%
commentstyle=\itshape\color{dkgreen},%
columns=[l]fullflexible,%
sensitive=true,%
morecomment=[s]{(*}{*)},%
moredelim=[is][\it\color{dkgreen}]{/*}{*/},
numberstyle=\color{gray},
stringstyle=\color{mauve},
emphstyle=[1]{\color{red}},
emph=[1]{Array, Algebriac, true, false},
escapeinside={*?}{?*},%
keepspaces=true,%
literate=%
{'}{$^\prime$}{1}%
 {=>}{$\Rightarrow$}{2}% -> vs => lol
 {<=}{$\le$}{1}%
 {>=}{$\ge$}{1}%
 {<>}{$\ne$}{1}%
 {=:=}{$\eqbi$}{1}%
 {/\\}{$\land$}{1}%
 {\\/}{ $\lor$ }{3}%
 {and}{$\&\&$}{2}%
 {\ or(}{ $\lor$(}{3}%
 {not\ }{$\lnot$ }{2}%
 {not(}{$\lnot$(}{2}%
 {+->}{\texttt{+->}}{3}%
 {+->}{$\mapsto$}{2}%
 {-->}{$\Longrightarrow$}{2}%
 {->}{$\Rightarrow$}{2}%
 {<-}{$\leftarrow$}{2}%
 {<->}{$\Leftrightarrow$}{2}%
 {<<}{$\subseteq$}{2}
 {iin}{$\in$}{2}
 {**}{$\union$}{2}
 {\^\^}{$\cap$}{2}
 {forall}{$\forall$}{2}
 {exists}{$\exists$}{2}
 {empty}{$\emptyset$}{2}
 {msigma}{$\sigma$}{1}
 {mtau}{$\tau$}{1}
 {mpi}{$\pi$}{1}
 {mrho}{$\rho$}{1}
 {_l}{$_{\ell}$}{2}
 {_r}{$_{r}$}{2}
}

\newcommand{\WhyRel}{WhyRel}

\newcommand{\Img}{\mbox{\large\textbf{`}}}
\newcommand{\rw}[1]{\keyw{rw}\,#1} % read + write abbrev
\renewcommand{\emptyset}{\varnothing}

\newcommand{\union}{\mathbin{\mbox{\small$\cup$}}}

\newcommand{\eqdef}{\mathrel{\,\hat{=}\,}}

\newcommand{\keyw}[1]{\ensuremath{\mathsf{#1}}} 

\newcommand{\sing}[1]{\{#1\}} % singleton region

\newcommand{\Rall}[1][]{\mathit{R}}

\newcommand{\syncbi}[1]{\lfloor #1 \rfloor}
\newcommand{\splitbi}[2]{(#1|#2)}
\def\rightharpoonupfill{$\mathsurround=0pt \mathord- \mkern-6mu
  \cleaders\hbox{$\mkern-2mu \mathord- \mkern-2mu$}\hfill
  \mkern-6mu \mathord\rightharpoonup$}

\def\leftharpoonupfill{$\mathsurround=0pt \mathord\leftharpoonup \mkern-6mu
  \cleaders\hbox{$\mkern-2mu \mathord- \mkern-2mu$}\hfill
  \mkern-6mu \mathord-$}

\def\overleftharpoonup#1{\vbox{\ialign{##\crcr
      \leftharpoonupfill\crcr\noalign{\kern-1pt\nointerlineskip}
      $\hfil\displaystyle{#1}\hfil$\crcr}}}

\def\overrightharpoonup#1{\vbox{\ialign{##\crcr
      \rightharpoonupfill\crcr\noalign{\kern-1pt\nointerlineskip}
      $\hfil\displaystyle{#1}\hfil$\crcr}}}

\def\overleftrightharpoonup#1{\vbox{\ialign{##\crcr
      \leftharpoonupfill\hspace*{-.7em}\rightharpoonupfill\crcr\noalign{\kern-1pt\nointerlineskip}
      $\hfil\displaystyle{#1}\hfil$\crcr}}}
\newcommand{\Left}[1]{\overleftharpoonup{#1}} % projections on various syntax; no longer formulas
\newcommand{\Right}[1]{\overrightharpoonup{#1}}
\newcommand{\Agr}{\ensuremath{\mathbb{A}}}
\newcommand{\eqbi}{\mathrel{\ddot{=}}}

\newcommand{\lexop}{\text{\color{blue}\textbf{$\langle\hspace*{-2.2pt}[$}}} % open left expression
\newcommand{\lexcl}{\text{\color{blue}\textbf{$\langle\hspace*{-2.3pt}]$}}} % close left expression
\newcommand{\rexop}{\text{\color{blue}\textbf{$[\hspace*{-2.4pt}\rangle$}}} % open right expression
\newcommand{\rexcl}{\text{\color{blue}\textbf{$]\hspace*{-2.2pt}\rangle$}}} % close right expression

\newcommand{\leftex}[1]{ \text{\small$\langle\hspace*{-2.2pt}[$} #1 \text{\small $\langle\hspace*{-.50ex}]$}
}
\newcommand{\rightex}[1]{ \text{\small$[\hspace*{-2.2pt}\rangle$} #1 \text{\small$]\hspace*{-2.2pt}\rangle$} 
}

\newcommand{\leftF}[1]{\leftex{#1}}
\newcommand{\rightF}[1]{\rightex{#1}}

\renewcommand{\phi}{\varphi}

\definecolor{light-gray}{gray}{0.88}
\definecolor{dark-gray}{gray}{0.25}
\newcommand{\mt}[1]{\ensuremath{\mathit{#1}}}
\newcommand{\etrans}[1]{\ensuremath{\mathcal{E}\llbracket #1 \rrbracket}}
\newcommand{\utrans}[1]{\ensuremath{\mathcal{U}\llbracket #1 \rrbracket}}
\newcommand{\btrans}[1]{\ensuremath{\mathcal{B}\llbracket #1 \rrbracket}}
\newcommand{\ftrans}[1]{\ensuremath{\mathcal{F}\llbracket #1 \rrbracket}}

\begin{document}
%
% \title{Contribution Title\thanks{Supported by organization x.}}
\title{The \WhyRel\ Prototype for Modular Relational Verification of Pointer Programs}
\titlerunning{The \WhyRel\ Prototype for Relational Verification}

% \author{Ramana Nagasamudram\inst{1}\orcidID{0000-0001-9979-1292} \and
% Anindya Banerjee\inst{2}\orcidID{0000-0001-9979-1292} \and
% David A. Naumann\inst{1}\orcidID{0000-0002-7634-6150}}
\author{Ramana Nagasamudram\inst{1} \and Anindya Banerjee\inst{2} \and David A. Naumann\inst{1}}
\authorrunning{R. Nagasamudram, A. Banerjee, and D. A. Naumann.}
\institute{Stevens Institute of Technology, Hoboken, USA\\
  \email{rnagasam,dnaumann@stevens.edu}
  \and
  IMDEA Software Institute, Madrid, Spain\\
  \email{anindya.banerjee@imdea.org}}
%
% \author{\vspace*{-5ex}}
% \institute{\vspace*{-5ex}}
\maketitle              % typeset the header of the contribution
\begin{abstract}
% The abstract should briefly summarize the contents of the paper in
  Verifying relations between programs arises as a task in various
  verification contexts such as optimizing transformations, relating new
  versions of programs with older versions (regression verification), and
  noninterference.  However, relational verification for programs acting on
  dynamically allocated mutable state is not well supported by existing tools,
  which provide a high level of automation at the cost of restricting the
  programs considered.  Auto-active tools, on the other hand, require more
  user interaction but enable verification of a broader class of programs.
  This article presents WhyRel, a tool for the auto-active verification of
  relational properties of pointer programs based on relational region logic.
  WhyRel is evaluated through verification case studies, relying on SMT
  solvers orchestrated by the Why3 platform on which it builds.  Case studies
  include establishing representation independence of ADTs, showing
  noninterference, and challenge problems from recent literature.
\end{abstract}

\section{Introduction}

Relational properties encompass conditional equivalence of programs (as in
regression verification~\cite{Strichman2008}), noninterference (in which a
program is related to itself via a low-indistinguishability relation), and
other requirements such as sensitivity~\cite{BartheTraceLogic19}.  The problem
we address concerns tooling for the modular verification of relational
properties of heap-manipulating programs, including programs that act on
differing data representations involving dynamically allocated pointer
structures.

Modular reasoning about pointer programs is enabled through local reasoning
using frame conditions, procedural abstraction (i.e., reasoning under
hypotheses about procedures a program invokes), and data abstraction,
requiring state-based encapsulation.  For establishing properties of ADTs such
as representation independence, encapsulation plays a crucial role, permitting
implementations to rely on invariants about private state hidden from clients.
Relational verification also involves a kind of compositionality, the
\emph{alignment} of intermediate execution steps, which enables use of simpler
relational invariants and specs (see
e.g.~\cite{TerauchiAiken,KovacsSF13,ShemerGSV19}).

We aim for auto-active verification~\cite{LeinoM10}, accessible to developers,
as promoted by tools such as Dafny and Why3.  Users are expected to provide
specifications, annotations such as loop invariants and assertions, and, for
relational verification, alignment hints.  The idea is to minimize or
eliminate the need for users to manually invoke tactics for proof search.

Automated inference of specs, loop invariants, or program alignments
facilitates automated verification, and is implemented in some tools.  But in
the current state of the art these techniques are restricted to specs and
invariants of limited forms (e.g., only linear arithmetic) and seldom support
dynamically allocated objects.  So inference is beyond the scope of this
paper.

What is in scope is use of strong encapsulation, to hide information in the
sense that method specs used by clients do not expose internal representation
details, and to enable verification of modular correctness of a client, in the
sense that its behavior is independent from internal representations.
Achieving strong encapsulation for pointer programs, without undue restriction
on data and control structure, is technically challenging.  Auto-active tools
rely on extensive axiomatization for the generation of \emph{verification
  conditions (VCs)}; for high assurance the VCs should be justified with
respect to a definitional operational semantics of programs and specs.

In this article, we describe \WhyRel, a prototype for auto-active verification
of relational properties of pointer programs.  Source programs are written in
an imperative language with support for shared mutable objects (but no
subtyping), dynamic allocation, and encapsulation.  The assertion language is
first-order and, for expressing relational properties, includes constructs
that relate values of variables and pointer structures between two programs.
\WhyRel\ is based on relational region logic~\cite{BNNN22}, a relational
extension of region logic~\cite{RegLogJrnI,RegLogJrnII}.  Region logic
provides a flexible approach to local reasoning through the use of
\emph{dynamic frame} conditions~\cite{KassiosFM} which capture
\emph{footprint\/}s of commands acting on the heap.  Verification involves
reasoning explicitly about regions of memory and changes to them as
computation proceeds; flexibility comes from being able to express notions
such as parthood and separation in the same first-order setting.

Encapsulation is specified using a kind of dynamic frame, called a
\emph{dynamic boundary}: a footprint that captures a module's internal
locations.  Enforcing encapsulation is then a matter of ensuring that clients
don't directly modify or update locations in a module's boundary.  There are
detailed soundness proofs for the relational logic~\cite{BNNN22}, of which our
prototype is a faithful implementation.

\WhyRel\ is built on top of the Why3 platform\footnote{The Why3 distribution
  can be found at: \url{https://why3.lri.fr/}.} for deductive program
verification which provides infrastructure for verifying programs written in
WhyML, a subset of ML~\cite{Why3} with support for ghost code and
nondeterministic choice.  The assertion language is a polymorphic first-order
logic extended with support for algebraic data types and recursively and
inductively defined predicates~\cite{Why3Logic}.
% Why3 provides support for user-defined axioms, lemmas and ghost code.
Why3 generates VCs for WhyML which can then be discharged using a wide array
of theorem provers, from interactive proof assistants such as Coq and
Isabelle, to first-order theorem provers and SMT solvers such as Vampire,
Alt-Ergo and Z3.

Primarily, \WhyRel\ is used as a front end to Why3.  Users provide programs,
specs, annotations, and for relational verification, relational specs and
alignment specified using a specialized syntax for product programs.  \WhyRel\
translates source programs into WhyML, performing significant encoding so as
to faithfully capture the heap model and fine-grained framing formalized in
relational region logic.  VCs pertinent to this logic are introduced as
intermediate assertions and lemmas for the user to establish.  Verification is
done using facilities provided by Why3 and the primary mode of interaction is
through an IDE for viewing and discharging verification conditions.

Our approach is evaluated through a number of case studies performed in
\WhyRel{}, for which we rely entirely on SMT solvers to discharge proof
obligations.  The primary contribution is the development of a tool for
relational verification of heap manipulating programs which has been applied
to challenging case studies.  Examples formalized demonstrate the
effectiveness of relational region logic for alignment, for expressing heap
relations, and for relational reasoning that exploits encapsulation.

\paragraph{Organization.} Sec.~\ref{sec:tour} highlights aspects of specifying
programs and relational properties in \WhyRel\ using a stack ADT example.
Sec.~\ref{sec:alignment} discusses examples of program alignment.
Sec.~\ref{sec:implementation} gives an overview of the design of \WhyRel\ and
Sec.~\ref{sec:evaluation} provides highlights on experience using the
tool. Sec.~\ref{sec:related-work} discusses related work and
Sec.~\ref{sec:conclusion} concludes.

\section{A tour of WhyRel}
\label{sec:tour}

\paragraph{Programs and specifications.}  \WhyRel\ provides a lightweight
module system to organize definitions, programs, and specs.  Developments are
structured into interfaces and modules that implement interfaces.  In
addition, for relational verification, \WhyRel\ introduces the notion of a
\emph{bimodule}, described later, to relate method implementations between two
(\emph{unary}) modules.

We'll walk through aspects of specification in \WhyRel\ using the \code{STACK}
interface shown in Fig.~\ref{fig:stack-interface}, which describes a stack of
boxed integers with push and pop operations.  The interface starts by
declaring global variables, \code{pool} and \code{capacity}, and
client-visible fields of the \code{Cell} and \code{Stack} classes.  Variable
\code{pool} has type \code{rgn}, where a \emph{region} is a set of references,
and is used to describe objects notionally owned by modules implementing the
stack interface; \code{capacity} has type \code{int} and describes an upper
bound on the size of a stack.  The \code{Cell} class for boxed integers is
declared with a single field, \code{val}, storing an \code{int}.  The
\code{Stack} class is declared with three fields: \code{rep} of type region
keeps track of objects used to represent the stack, \code{size} of type
\code{int} stores the number of elements in the stack, and the ghost field
\code{abs} of type \code{intlist} (list of mathematical integers) keeps track
of an abstraction of the stack, used in specs.
% \code{rep} of type region to
% keep track of objects used to represent the stack, \code{size} of type
% \code{int} to store the number of elements in the stack, and a ghost field
% \code{abs} to keep track of a mathematical abstraction of the stack, used in
% specs.
Class definitions can be refined later by modules implementing the
interface: e.g., a module using a linked-list implementation might extend the
\code{Stack} class with a field \code{head} storing a reference to the list.
\begin{figure}[t]
    \begin{lstlisting}[language=whyrel]
interface STACK =
  public pool:rgn /* rgn: a set of references */  public capacity:int
  class Cell {val:int}  class Stack {rep:rgn; size:int; ghost abs:intlist}

  /* encapsulated locations */  
  boundary {capacity, pool, pool`any, pool`rep`any}

  public invariant stkPub = forall s: Stack iin pool. 0 <= s.size <= capacity
    /\ (forall t: Stack iin pool. s <> t -> s.rep ^^ t.rep << {null}) /\ ...

  meth Cell(self: Cell) : unit ...   meth getVal(self: Cell) : int ...
  meth Stack(self: Stack) : unit ensures {self iin pool} ...

  meth push(self: Stack, k: int) : unit
    requires {self iin pool /\ self.size < capacity}
    ensures  {self.abs = cons(k,old(self.abs)) /\ ...}
    /* allowed heap effects of implementations */
    effects  {rw {self}`any, self.rep`any, alloc; rd self,capacity}

  meth pop(self: Stack) : Cell
    requires {self iin pool /\ self.size > 0}
    ensures  {self.size = old(self.size)-1}
    ensures  {result.val = hd(self.abs) /\ self.abs = tl(old(self.abs))}
    \end{lstlisting}
  \vspace*{-3ex}\caption{\WhyRel\ interface for the Stack ADT}
  \label{fig:stack-interface}
\end{figure}

Heap encapsulation is supported at the granularity of modules through the use
of \emph{dynamic module boundaries} which describe locations internal to a
module.  A \emph{location} is either a variable or a heap location $o.f$,
where $o$ is an object reference and $f$ is its field.  In \WhyRel, module
boundaries are specified in interfaces and clients are enforced to not
directly read or write locations described by the boundary except through the
use of module methods.  For our stack example, the dynamic boundary is
\code{capacity, pool, pool`any, pool`rep`any}; expressed using image
expressions and the \code{any} datagroup.  Given a region $G$ and a field $f$
of class type, the \emph{image expression} $G\Img f$ denotes the region
containing the locations $o.f$ of all non-null references $o$ in $G$, where
$f$ is a valid field of $o$.  If $f$ is of type region, $G\Img f$ is the union
of the collection of reference sets $o.f$ for all $o$ in $G$.  For $f$ of
primitive type, such as \code{int} or \code{intlist}, $G\Img f$ is the empty
region.  The \emph{datagroup} \code{any} is used to abstract from concrete
field names: the expression \code{pool`any} is syntactic sugar for
\code{pool`val},\ldots,\code{pool`abs}.  Intuitively, the dynamic boundary in
Fig.~\ref{fig:stack-interface} says that clients may not directly read or
write \code{capacity}, \code{pool}, any fields of objects in \code{pool}, and
any fields of objects in the \code{rep} of any \code{Stack} in \code{pool}.

While encapsulation is specified at the level of modules, separation or
locality at finer granularities can be specified using module invariants.  The
stack interface defines a public invariant \code{stkPub} which asserts that
the \code{rep} fields of all \code{Stack} objects in \code{pool} are disjoint.
This idiom can be used to ensure that modifying one object has no effect on
any locations in the representation of another.  Clients can rely on public
invariants during verification, but modules implementing the interface must
ensure they are preserved by module methods.  Additionally, modules may define
private invariants that capture conditions on internal state; provided these
refer only to encapsulated locations, i.e., the designated boundary
\emph{frames} these invariants, clients are exempt from reasoning about
them~\cite{Hoare:data:rep}.

Finally, the \code{STACK} interface defines specs for initializers (methods
\code{Cell} and \code{Stack}) and public specs for client-visible methods
\code{getVal}, \code{push}, and \code{pop}.  Notice that the stack initializer
ensures \code{self} is added to the boundary (through post \code{self iin
  pool}) and stack operations require \code{self} to be part of the boundary
(through pre \code{self iin pool}).  Specs for push and pop are standard,
using ``old'' expressions to precisely capture field updates.  \WhyRel's
assertion language is first-order and includes constructs such as the
points-to assertion $x.f = e$ and operations on regions such as subset and
membership.  In addition to pre- and post-conditions, each method is annotated
with a frame condition in an \code{effects} clause that serves to constrain
heap effects of implementations.  Allowable effects are expressed using
read/write (\code{rw}) or read (\code{rd}) of locations or location sets,
described by regions.  For example, the \code{effects} clause for \code{push}
says that implementations may read/write any field of \code{self} and any
field of any objects in \code{self.rep}.  The distinguished variable
\code{alloc} is used to indicate that \code{push} may dynamically allocate
objects.

In our development, we build two modules that implement the interface in
Fig.~\ref{fig:stack-interface}: one using arrays, \code{ArrayStack} and
another using linked-lists, \code{ListStack}.  Both rely on private invariants
on encapsulated state that capture constraints on their pointer
representations and its relation to \code{abs}, the mathematical abstraction
of stack objects.
The private invariant of \code{ListStack}, for example, says that \code{Cell}
values in the linked-list of any \code{Stack} in \code{pool} are in
correspondence with values stored in \code{abs}.
% For instance, the private invariant of \code{ListStack}
% says that \code{Cell} values in the linked-list of any \code{Stack} in
% \code{pool} are in a one-to-one correspondence with values stored in
% \code{abs}.

% TODO highlight rw and rd as keywords -- in blue
\begin{figure}[t]
  \begin{lstlisting}[language=whyrel]
module Client =
  meth prog (n: int) : int
    requires { 0 <= n < capacity /\ ... }
    effects  { rw alloc, pool, pool`any, pool`rep`any; rd n, capacity }
  = var i: int in var c: Cell in
    var stk: Stack in stk := new Stack; Stack(stk);
    while (i < n) do push(stk,i); i:=i+1 done; i := 0;
    while (i < n) do c:=pop(stk); result:=result+getVal(c); i:=i+1 done;

  meth prog (n: int|n: int) : (int|int)  /* Relational spec for prog */
    requires { n =:= n /\ Both(0 <= n < capacity /\ ... ) }
    ensures  { result =:= result }
  \end{lstlisting}
    \vspace*{-3ex}\caption{Example client for \code{STACK} and relational spec for
    equivalence}
  \label{fig:stack-client}
\end{figure}
\paragraph{Example client, equivalence spec, and verification.}  We now turn
attention to an example client, \code{prog}, shown in
Fig.~\ref{fig:stack-client}.  This program computes the sum
$\Sigma_{i=0}^{n} i$, albeit in a roundabout fashion, using a stack.  The
frame condition of \code{prog} mentions the boundary for \code{STACK}, but
this is fine since the client respects \WhyRel's encapsulation discipline,
modifying encapsulated locations solely through calls to methods declared in
the \code{STACK} interface.
For this client, our goal is to establish equivalence when linked against
either implementations of \code{STACK}.  Let the \emph{left} program be the
client linked against \code{ArrayStack}, and the \emph{right} the client linked
against \code{ListStack}% \footnote{For things that come in twos, we'll use the
  % terms \emph{left} and \emph{right}: e.g., for the
  % two programs being related, the two states they operate on, the two heap
  % effects they describe, etc.}.
Equivalence is expressed using the relational
spec shown in Fig.~\ref{fig:stack-client}.
For brevity, we omit frame conditions when describing relational specs.

This relational spec relates two versions of \code{prog}; the notation
\code{(n:int | n:int)} is used to declare that both versions expect \code{n}
as argument.  The pre-relation requires equality of inputs: \code{n =:= n}
says that the value of \code{n} on the left is equal to the value of \code{n}
on the right.  We use (\code{=:=}), instead of ($=$) to distinguish between
values on the left and the right\footnote{Note in particular that $x \eqbi y$
  is not the same as $y \eqbi x$}.  The relational spec requires the two
states being related to satisfy the unary precondition for the client, as
indicated by \code{Both(...)}.  The post-relation, \code{result =:= result},
asserts equality on returned values.  In \WhyRel, relational specs capture a
$\forall\forall$ termination-insensitive property: \emph{terminating
  executions of the programs being related, when started in states related by
  the pre-relation, will result in states related by the post-relation}.
% Relational specs in \WhyRel\ capture a
% $\forall\forall$ termination-insensitive property: for any two states that are
% related by the pre-relation, if executions of the programs considered
% terminate in two post-states, they will be related by the post-relation.

\WhyRel\ supports two approaches to verifying relational properties.  The
first reduces to proving functional properties of the programs involved.  For
instance, equivalence of the client when linked against the two stack
implementations is immediate if we prove that \code{prog} indeed computes the
sum of the first \code{n} nonnegative integers.

However, this approach neither lends well to more complicated programs and
relational properties, nor does it allow us to exploit similarities between
related programs or reason modularly using relational specs.  The alternative
is to prove the relational property using a convenient alignment of the two
programs.  Alignments are represented syntactically in \WhyRel\ using
\emph{biprograms} which pair points of interest between two programs so that
their effects can be reasoned about in tandem.  If the chosen alignment is
\emph{adequate} in the sense of capturing all pairs of executions of the
related programs, relational properties of the alignment entail the
corresponding relation between the underlying programs.
\begin{figure}[t]
  \begin{lstlisting}[language=whyrel]
meth prog (n: int | n: int) : (int | int)
= var i: int | i: int in var c: Cell | c: Cell in
  var stk: Stack in |_ stk := new Stack _|; |_ Stack(stk) _|;
  while (i < n) | (i < n) do |_ push(stk,i) _|; |_ i:=i+1 _| done; |_ i:=0 _|;
  while (i < n) | (i < n) do |_ c:=pop(stk) _|;
    |_ result:=result+getVal(c) _|; |_ i:=i+1 _| done;
\end{lstlisting}
    \vspace*{-2ex}\caption{Alignment for example stack client}
  \label{fig:client-biprogram}
\end{figure}

The biprogram for \code{prog} is shown in Fig.~\ref{fig:client-biprogram}.
The alignment it captures is maximal: every control point in one version of
the client is paired with itself in the other version.  The construct
$\splitbi{C}{C'}$ pairs a command $C$ on the left with a command $C'$ on the
right, and the \emph{sync} form $\syncbi{C}$ is syntactic sugar for
$\splitbi{C}{C}$; e.g., the biprogram for \code{prog} aligns the two
allocations using \code{|_stk := new Stack_|}.  Further, this biprogram aligns
both loops in \emph{lockstep}, indicated using the syntax \code{while e|e' do
  ... done}.  This alignment pairs a loop iteration on the left with a loop
iteration on the right and requires the loop guards be in agreement: here,
that \code{i < n} on the left is true just when \code{i < n} on the right is.
Calls to stack operations are aligned in the loop body using the sync
construct to facilitate modular verification of relational properties by
indicating that relational specs for \code{push} and \code{pop} are to be
used.

To prove the spec (in Fig.~\ref{fig:stack-client}) about the biprogram in
Fig.~\ref{fig:client-biprogram} we reason as follows: after allocation
\code{stk} on both sides is initialized to be the empty stack.  The first
lockstep aligned loop which pushes integers from $0,\ldots,\code{n}$ maintains
as invariant equality on \code{i} and on the mathematical abstractions the two
stacks represent, i.e., \code{i =:= i /\\ stk.abs =:= stk.abs}.  The second
lockstep aligned loop which pops the stacks and increments \code{result}
maintains as invariant agreement on the stack abstractions and \code{result},
the key conjunct being \code{result =:= result}.  This is sufficient to
establish the desired post-relation.  Importantly, the loop invariants are
simple to prove---they only contain equalities between variables---and we
don't have to reason about the exact contents of the two stacks involved.

\paragraph{Relational specs for Stack and verification.}  The reasoning
described above relies on knowing the method implementations in
\code{ArrayStack} and \code{ListStack} are equivalent.  We need relational
specs for push which state that given related inputs, the contents represented
by the two stacks are the same; and for pop, which state that given related
inputs, the values of the returned \code{Cell}s are the same.
\begin{figure}[t]
  \begin{lstlisting}[language=whyrel]
bimodule REL_STACK (ArrayStack | ListStack) =
  coupling stackCoupling = forall s: Stack iin pool | s: Stack iin pool.
    s =:= s -> s.abs =:= s.abs /\ ...

  meth Stack(self: Stack | self: Stack) : (unit | unit)
    ensures {self =:= self /\ ...} = /* biprogram for Stack */
  meth push(self: Stack | self: Stack) : (unit | unit)
    requires {self =:= self /\ ... }
    ensures  {self.abs =:= self.abs /\ ... } = /* biprogram for push */
  meth pop(self:Stack | self:Stack) : (Cell | Cell)
    requires {self =:= self /\ Both (self iin pool) /\ Both (self.size > 0)}
    ensures  {... /\ result.val =:= result.val} = /* biprogram for pop */
\end{lstlisting}
\vspace*{-2ex}
  \caption{Bimodule for Stack; excerpts}
  \label{fig:stack-bimodule}
\end{figure}

Fig.~\ref{fig:stack-bimodule} shows a bimodule, \code{REL_STACK}, relating the
two implementations of \code{STACK}.  It includes relational specs for the
stack operations along with biprograms used for verification.  The bimodule
maintains a \emph{coupling} relation which relates data representations used
by the two stack implementations.  Concretely, the coupling here states that
related stacks in \code{pool} represent the same abstraction.  Note that
quantifiers in relation formulas bind pairs of variables; and the equality
\code{s =:= s} in \code{stackCoupling} is not strict pointer
equality, but indicates correspondence.
Strict pointer equality is too strong as it would not allow for modeling
allocation as a nondeterministic operation or permit differing allocation
patterns between programs being related.  Behind the scenes, \WhyRel\
maintains a partial bijection $\pi$ between allocated references in the two
states being related.  The relation \code{x =:= y}, where \code{x} and
\code{y} are pointers, states that \code{x} in the left state is in
correspondence with \code{y} in the right state w.r.t $\pi$, i.e.,
$\pi(\code{x}) = \code{y}$.

The relational spec for the initializer \code{Stack} ensures \code{self =:=
  self}, which is required in the specs for \code{push} and \code{pop}.  Like
other invariants, coupling relations are meant to be framed by the boundary
and are required to be preserved by module methods being related.
Encapsulation allows for coupling relations to be hidden so that clients are
exempt from reasoning about them.

The steps taken to complete the Stack development and verify equivalence of
two versions of its client are as follows: (i) build the \code{STACK}
interface in \WhyRel{}, with public invariants clients can rely on and a
boundary that designates encapsulated locations; (ii) develop two modules
refining this interface, \code{ArrayStack} and \code{ListStack}, and verify
that their implementations conform to \code{STACK} interface specs, relying on
any private invariants that capture conditions on encapsulated state; (iii)
provide a bimodule relating the two stack modules and prove equivalence of
stack operations, relying on a coupling relation that captures relationships
between pointer structures used by the two modules; (iv) verify the client
with respect to specs given in \code{STACK} and prove it respects \WhyRel's
encapsulation regime; and finally (v) develop a bimodule for the client and
verify equivalence using relational specs for stack methods.

\section{Patterns of alignment}
\label{sec:alignment}

Well chosen alignments help decompose relational verification, allowing for
the use of simple relational assertions and loop invariants.  In this section,
we'll look at examples of biprograms that capture alignments that aren't
maximal, unlike the \code{STACK} client example in Sec.~\ref{sec:tour}.  We
don't formalize the syntax of biprograms here, but we show representative
examples.  When discussing examples, we'll omit frame conditions and other
aspects orthogonal to alignment.

\begin{figure}[t]
  \centering
  \begin{subfigure}{.55\textwidth}
    \begin{lstlisting}[language=whyrel]
meth mult(n: int, m: int) =
  i := 0;
  while (i < n) do j:=0;
    while (j < m) do
      result := result+1; j := j+1
    done; i := i+1 done;
    \end{lstlisting}
  \end{subfigure}
  \begin{subfigure}{.35\textwidth}
    \begin{lstlisting}[language=whyrel]
meth mult(n:int, m:int) =
  i := 0;
  while (i < n) do
    result := result+m;
    i := i+1
  done;
    \end{lstlisting}
  \end{subfigure}
\vspace*{-3ex}
  \caption{Two versions of a simple multiplication routine}
  \label{fig:mult-unary}
\end{figure}
\paragraph{Differing control structures.}
Churchill et al.~\cite{ChurchillP0A19} develop a technique for proving
equivalence of programs using state-dependent alignments of program traces.
They identify a challenging problem for equivalence checking, shown in
Fig.~\ref{fig:mult-unary}, which compares two procedures for multiplication
with different control flow.  For automated approaches to relational
verification, their example is challenging because of the need to align an
unbounded number $m$ of loop iterations on the left with a single iteration on
the right.

To prove equivalence, we verify the biprogram shown in
Fig.~\ref{fig:mult-biprogram} with respect to a relational spec with
pre-relation \code{n =:= n /\\ m =:= m} and post-relation \code{result =:=
  result}; i.e., agreement on inputs results in agreement of outputs.  Unlike
the stack client biprogram shown in Fig.~\ref{fig:client-biprogram}, the
alignment embodied here is not maximal---indeed, such alignment would not be
possible due to the differing control structure.  Similarities are still
exploited by aligning the outer loops in lockstep and the left inner loop with
the assignment to \code{result} on the right.
\begin{figure}[t]
\begin{lstlisting}[language=whyrel]
meth mult(n: int, m: int | n: int, m: int) : (int | int) =
  |_ i := 0 _|;
  while (i < n) | (i < n) do  invariant { i =:= i /\ result =:= result }
    ( j := 0; while (j < m) do result := result+1; j := j+1 done
    | result := result+m ); 
    assert { *<|result = old(result)+m*<] };
    |_ i := i+1 _| done;
\end{lstlisting}
\vspace*{-2ex}
\caption{Biprogram for example in Fig.~\ref{fig:mult-unary}}
\label{fig:mult-biprogram}
\end{figure}

A simple relational loop invariant which asserts agreement on \code{i} and
\code{result} is sufficient for proving equivalence.  To show this is
invariant, we need to establish that the inner loop on the left has the effect
of incrementing \code{result} by \code{m}, thereby maintaining equality on
\code{result} after the inner loop.  In Fig.~\ref{fig:mult-biprogram} this is
indicated by the assertion after the left inner loop.  The notation
$\leftF{P}$ (resp. $\rightF{P}$) is used to state that the unary formula $P$
holds in the left (and resp. right) state.

% \code{*<| P *<]} is
% used to assert that the unary formula \code{P} holds on the left, and ``old''
% expressions are used to capture the values of variables at the start of an
% outer loop iteration.
\begin{figure}[t]
  \begin{subfigure}{.45\textwidth}
    \begin{lstlisting}[language=whyrel]
meth sumpub (l: List) : int =
  p:=l.head; s:=0;
  while (p <> null) do
    if p.pub then
      s:=s+p.val
    end;
    p:=p.nxt
  done;
  result:=s;    
    \end{lstlisting}
  \end{subfigure}
  \begin{subfigure}{.45\textwidth}
    \begin{lstlisting}[language=whyrel]
meth sumpub (l: List | l: List) : int =
  |_ p:=l.head _|; |_ s:=0 _|;
  while (p <> null) | (p <> null) .
    *<| not p.pub *<] | [> not p.pub |> do
    ( if p.pub then s:=s+p.val end;
      p:=p.nxt
    | if p.pub then s:=s+p.val end;
      p:=p.nxt)
  done; |_ result:=s _|;    
    \end{lstlisting}
  \end{subfigure}
\vspace*{-3ex}
  \caption{Summing up public elements of a linked list: program and alignment}
  \label{fig:sumpub}
\end{figure}

\paragraph{Conditionally aligned loops.} Examples so far have concerned
lockstep aligned loops, requiring a one-to-one correspondence between loop
iterations.  However, this condition is often too restrictive.  \WhyRel\
provides for other patterns of loop alignment, including those that account
for conditions on data values.  Consider for example the program shown in
Fig.~\ref{fig:sumpub} which traverses a linked list and computes the sum of
all elements marked public, indicated in each element's \code{pub} field.  The
program satisfies the following noninterference property, with relational spec:
\begin{lstlisting}[language=whyrel]
meth sumpub(l: List | l: List) : (int | int)
  requires { Both(listpub(l,xs)) /\ xs =:= xs }
  ensures { result =:= result }
\end{lstlisting}
Here \code{listpub(l,xs)} is a predicate which asserts that the
\emph{sequence} of public values reachable from the list pointer \code{l} is
realized in \code{xs}, a mathematical list of integers.  Intuitively, this
specification captures the property that the result of \code{sumpub} does not
depend on the values of nonpublic elements in the input list \code{l}.
Showing the program computes exactly the sum of public elements: \code{result
  = sum(xs)} would imply the desired noninterference property.  However, to
showcase support \WhyRel\ offers for non-lockstep alignments, we'll establish
noninterference by conditionally aligning the loops in the two copies of
\code{sumpub} (see Fig.~\ref{fig:sumpub}).

The alignment is as follows: if \code{p} is a nonpublic node on one side,
perform a loop iteration on that side, pausing the iteration on the other; and
if \code{p} on both sides is public, perform lockstep iterations of both
loops.  This has the effect of incrementing \code{s} exactly when both sides
are visiting public nodes, the values of which are guaranteed to be the same
by the relational precondition.  The biprogram expresses this alignment
through the use of additional annotations, called \emph{alignment guards}
which are general relation formulas and express conditions that lead to
left-only, right-only, or lockstep iterations.  The left alignment guard
\code{*<|not p.pub*<]} indicates that left-only loop iterations are to be
performed when \code{p} on the left is not public.  The right alignment guard
expresses a similar condition when \code{p} on the right is not public.
Iterations proceed in lockstep when both alignment guards are false, i.e.,
when \code{Both(p.pub)} is true.

This biprogram maintains \code{exists xs|xs. Both(listpub(p,xs)) /\\ xs=:=xs
  /\\ s=:=s} as loop invariant, which implies the desired post-relation.  This
invariant states that \code{p} on both sides points to the same sequence of
public values as captured by \code{listpub(p,xs)} and that there is agreement
on the sum \code{s} computed so far.  During verification, we must establish
that left-only, right-only, and lockstep iterations of the aligned loops
preserve this invariant.  Due to the alignment, the value of \code{s} is only
updated during lockstep iterations and its straightforward to show
preservation.  For one-sided iterations, reasoning relies on knowing that the
sequence of public values pointed to by \code{p} remains the same.

\section{Encoding and design}
\label{sec:implementation}

We implement \WhyRel\ in OCaml, relying on a library provided by Why3 for
constructing WhyML parse trees.  Source programs are parsed and typechecked
before being translated to WhyML.  Prior to translation, \WhyRel\ performs a
variety of checks and transformations: primary among these is a check that
clients respect encapsulation and that any biprograms provided by users are
adequate.  Proof obligations pertinent to relational region logic are
generated in the form of intermediate assertions in WhyML programs and lemmas
for the user to prove.  In this section, we provide an overview of some
aspects of our implementation, focusing on the translation to WhyML.

\paragraph{Encoding program states.} References are represented using an
abstract WhyML type \code{reference} with a distinguished element,
\code{null}.  The only operation supported on reference values is equality;
\WhyRel\ does not deal with pointer-arithmetic.  Regions are encoded as ghost
state, using a library for mathematical sets provided by Why3.  Set operations
on regions are inherently supported, and we axiomatize image expressions: for
each field $f$, \WhyRel\ generates a Why3 function symbol \code{img_f} along
with an axiom that captures the meaning of $G\Img f$.

Program states are encoded using WhyML records.  An example is shown in
Fig.~\ref{fig:state-encoding}.  The \code{state} type includes at least two
mutable components called \code{alloct} and \code{heap}.  The component
\code{alloct} stores a map from references to object types and keeps track of
allocated objects; \code{heap} is itself a record with one mutable component
per field in the source program that stores a map from references to values.
The set of values includes references, Why3 mathematical types such as arrays
and lists, regions, and primitive types such as \code{int} and \code{bool}.
In addition, the \code{state} type contains one mutable field per global
variable in the source program, storing a value of the appropriate type.  The
\code{state} type is annotated with a WhyML invariant that captures
well-formedness.  This invariant includes conditions such as \code{null} never
being allocated, no dangling references, and typing constraints: for example,
the \code{nxt} field of a \code{Node} is itself a \code{Node}.
\begin{figure}[t]
  \begin{subfigure}{0.4\textwidth}
    \begin{lstlisting}[language=whyrel]
/* class defs */
class Cell {
  val: int;
  ghost rep: rgn; }

class Node {
  curr: Cell;
  nxt: Node; }

/* global vars */
public pool : rgn
    \end{lstlisting}
  \end{subfigure}
  \begin{subfigure}{0.55\textwidth}
    \begin{lstlisting}[language=why3]
type reftype = Cell | Node (*class names*)
type heap = {
  mutable val: map reference int;
  mutable ghost rep : map reference Rgn.t;
  mutable curr: map reference reference;
  mutable nxt: map reference reference }
type state = {
  mutable alloct: map reference reftype;
  mutable heap: heap;
  mutable ghost pool: rgn }
invariant {not(Map.mem null alloct) /\ *?\ldots?*}

(* axiomatization of G`nxt *)
function img_nxt : state -> Rgn.t -> Rgn.t
axiom img_nxt_ax : forall s, r, p.
  Rgn.mem p (img_nxt s r) <-> exists q.
      s.alloct[q] = Node /\ Rgn.mem q r
    /\ p = s.head.nxt[q]
\end{lstlisting}
  \end{subfigure}
\vspace*{-3ex}
  \caption{State encoding: \WhyRel\ source on left, encoding in WhyML on right.}
  \label{fig:state-encoding}
\end{figure}

\paragraph{Translating unary programs and effects.} \WhyRel\ translates unary
programs into WhyML functions that act on our encoding of states.  Commands
that modify the heap are modeled as updates to an explicit state parameter,
and local variables, parameters, and the distinguished \code{result} variable
are encoded using WhyML reference cells.  Object parameters are modeled using
the \code{reference} type and a typing assumption.  Translation of control
flow statements is straightforward.  For programs with loops, \WhyRel\
additionally adds a diverges clause to the generated WhyML function: this
indicates that the function may potentially diverge, avoiding generation of
VCs for proving termination.  While Why3 supports reasoning about total
correctness, we're only concerned with partial correctness.
Fig.~\ref{fig:unary-trans} shows an example translation.
\begin{figure}[t]
  \begin{subfigure}{0.45\textwidth}
\begin{lstlisting}[language=whyrel]
meth m (c: Cell, i: int) : int
  requires { c.val >= 0 }
= while (i >= 0) do
    invariant { c.val >= 0 }
    c.val := c.val+i;
    i := i-1
  done;
  result := c.val
\end{lstlisting}
 \end{subfigure}
 \begin{subfigure}{0.45\textwidth}
\begin{lstlisting}[language=why3]
let m (s:state) (c:reference) (i:int)
  : int diverges
  requires { s.alloct[c] = Cell }
  requires { s.heap.val[c] >= 0 }
= let result = ref 0 in
  let c = ref c in
  let i = ref i in
  while (!i >= 0) do
    invariant { s.heap.val[!c] >= 0 }
    (* c.val := c.val + i *)
    s.heap.val <- Map.add !c
      (s.heap.val[!c]+!i) s.heap.val;
    i := !i-1
  done;
  result := s.heap.val[!c]; !result
\end{lstlisting}
  \end{subfigure}
\vspace*{-2ex}
  \caption{Program translation example: \WhyRel\ program on the left, WhyML
    translation on the right; frame conditions omitted.}
  \label{fig:unary-trans}
\end{figure}

Translation of frame conditions requires care given our encoding of states.
As an example, the writes for method $\code{m}$ shown in Fig.~\ref{fig:unary-trans}
would include $\rw \sing{\code{c}}\Img \code{val}$ due to the write to, and
read of, field \code{val} of object \code{c}.  Correspondingly, in the Why3
translation, component \code{val} of \code{s.heap} is updated; so specifying
the function in Why3 requires adding \code{writes \{s.heap.val\}} as
annotation.  However, this isn't the granularity we want since it implies the
field \code{val} of any reference can be written.  Hence, \WhyRel\ generates
an additional postcondition for method \code{m}: \code{wr_framed_val (old s) s
  (Rgn.singleton c)}, where
\begin{lstlisting}[language=whyrel]
predicate wr_framed_val (s: state) (t: state) (r: rgn) = forall p: reference.
  s.alloct[p] = Cell /\ p *?$ \notin $?* r -> s.heap.val[p] = t.heap.val[p]
\end{lstlisting}
With this postcondition, callers of \code{m} (in WhyML) can rely on knowing
that the \code{val} fields of only references in $\sing{\code{c}}$ are
modified.

\paragraph{Biprograms.} \WhyRel\ translates biprograms into product programs;
specifically, WhyML functions that act on a pair of states\footnote{In
  reality, generated WhyML functions act on a pair of states and a bijective
  renaming of references allocated in these states.  This is to cater for
  relation formulas such as $x \eqbi y$ where $x$ and $y$ are references.
  However, this additional parameter is not important to our discussion here,
  so we avoid mentioning it.}.  Before translation, it performs an adequacy
check to ensure the biprogram is well-formed.  Recall that adequacy here means
that all computations of the underlying unary programs are covered by their
aligned biprogram.  Adequacy ensures that a relational judgment about the
biprogram entails the expected relation between the underlying unary programs.
The check \WhyRel\ performs is syntactic and defined using projection
operations on biprograms.  Given a biprogram $CC$, the left projection
$\Left{CC}$ (and resp. the right projection $\Right{CC}$) extracts the unary
program on the left (and resp. the right).  As an example, the left projection
of \code{|_c.f:=g_|; (x:=c.f | skip)} is \code{c.f:=g; x:=c.f} and its right
projection is \code{c.f:=g}.  For adequacy, given unary programs $C$ and $C'$
and their aligned biprogram $CC$, it suffices to check whether
$\Left{CC} \equiv C$ and $\Right{CC} \equiv C'$~\cite{BNNN22}.
\begin{figure}[t]
  \[\begin{array}{lll}
% bisplits      
     \btrans{\code{C|C'}}(\Gamma_l, \Gamma_r) & \eqdef &
        \utrans{\code{C}}(\Gamma_l) ;\; \utrans{\code{C'}}(\Gamma_r) \\
% bimethod calls
       \btrans{\code{|_m(x|y)_|}}(\Gamma_l, \Gamma_r) & \eqdef &
         \mt{apply} (\Phi(\code{m}), [\Gamma_l.\code{st}; \Gamma_r.\code{st}; \etrans{\code{x}}(\Gamma_l); \etrans{\code{y}}(\Gamma_r)] ) \\
% bisyncs
      \btrans{\code{|_C_|}}(\Gamma_l,\Gamma_r) & \eqdef &
        \btrans{\code{C|C}}(\Gamma_l,\Gamma_r) \\
% sequence
      \btrans{\code{C; C'}}(\Gamma_l, \Gamma_r) & \eqdef &
        \btrans{\code{C}}(\Gamma_l, \Gamma_r) ;\; \btrans{\code{C'}}(\Gamma_l, \Gamma_r) \\
% variable blocks
      \btrans{\code{var \x:T|x:T' in CC}}
         (\Gamma_l,\Gamma_r) & \eqdef &
          \code{let} \; x_l  = \code{def(T)} \; \code{in let} % \\
 \; x_r  = \code{def(T')} \; \code{in} \\
 & &       \btrans{\code{CC}}([\Gamma_l \mid \code{x} :\, x_l],[\Gamma_r \mid \code{x} : \, x_r]) \\
% aligned conditionals
      \btrans{\code{if E|E' then CC else DD}}(\Gamma_l,\Gamma_r) & \eqdef &
        \code{assert} \; \{ \etrans{\code{E}}(\Gamma_l)
                = \etrans{\code{E'}}(\Gamma_r) \} ; \\
  & &    \code{if} \; \etrans{\code{E}}(\Gamma_l) \;
\code{then} \; \btrans{\code{CC}} \; \code{else} \; \btrans{\code{DD}} \\
% loops with trivial alignment guards      
      \btrans{\code{while E|E' do DD}}(\Gamma_l,\Gamma_r) & \eqdef &
      \code{while} \; \etrans{\code{E}}(\Gamma_l) \; \code{do} \\
& & \;\; \code{invariant} \; \{ \etrans{\code{E}}(\Gamma_l) = \etrans{\code{E'}}(\Gamma_r) \} \\
& & \;\; \btrans{\code{CC}}(\Gamma_l,\Gamma_r) \\
% conditionally aligned loops
%       \btrans{\code{while E|E'. aP|aPP do DD}}
%          (\Gamma_l, \Gamma_r) & \eqdef &
%          \code{while} \; (\etrans{\code{E}}(\Gamma_l) \lor \etrans{\code{E'}}(\Gamma_r)) \; \code{do}
% \; \code{inv} \; \{ \mathcal{A} \} \\
% & &       \;\;\;\; \code{if} \; (\etrans{\code{E}}(\Gamma_l) \land \ftrans{\mathcal{P}}(\Gamma_l,\Gamma_r)) \\
% & &       \;\;\;\; \code{then} \; \utrans{\Left{\code{CC}}}(\Gamma_l) \\
% & &       \;\;\;\; \code{else if} \; (\etrans{\code{E'}}(\Gamma_r) \land \ftrans{\mathcal{P'}}(\Gamma_l,\Gamma_r)) \\
% & &       \;\;\;\; \code{then}\; \utrans{\Right{\code{CC}}}(\Gamma_r) \\
% & &       \;\;\;\; \code{else}\; \btrans{\code{CC}}(\Gamma_l,\Gamma_r) \\
% & & \\
%       \mbox{where } \mathcal{A} & \equiv &
%             \begin{array}{ll}
%             &  (\etrans{E}(\Gamma_l) \land \ftrans{\mathcal{P}}(\Gamma_l,\Gamma_r)) \\
%       \lor  & (\etrans{E'}(\Gamma_r) \land \ftrans{\mathcal{P'}}(\Gamma_l,\Gamma_r)) \\
%       \lor  & (\lnot \etrans{E}(\Gamma_l) \land \lnot \etrans{E'}(\Gamma_r)) \\
%        \lor & (\etrans{E}(\Gamma_l) \land \etrans{E'}(\Gamma_r))
                %               \end{array}
 \btrans{\code{while E|E'. aP|aPP do DD}}
      (\Gamma_l, \Gamma_r) & \eqdef  \\
    \end{array}\] \vspace{-2ex}
\[  \begin{array}{lll}
% conditionally aligned loops
      % & \btrans{\code{while E|E'. aP|aPP do DD}}
      % (\Gamma_l, \Gamma_r) \eqdef  \\
& \;\;\;\; \code{while} \; (\etrans{\code{E}}(\Gamma_l) \lor \etrans{\code{E'}}(\Gamma_r)) \; \code{do}
\; \code{invariant} \; \{ \mathcal{A} \} \\
&\;\;\;\;       \;\;\;\; \code{if} \; (\etrans{\code{E}}(\Gamma_l) \land \ftrans{\mathcal{P}}(\Gamma_l,\Gamma_r))
\; \code{then} \; \utrans{\Left{\code{CC}}}(\Gamma_l) \\
&\;\;\;\;       \;\;\;\; \code{else if} \; (\etrans{\code{E'}}(\Gamma_r) \land \ftrans{\mathcal{P'}}(\Gamma_l,\Gamma_r))
\; \code{then}\; \utrans{\Right{\code{CC}}}(\Gamma_r)
  \; \code{else}\; \btrans{\code{CC}}(\Gamma_l,\Gamma_r) \\[1ex]
& \;\;\; \mbox{where } \mathcal{A} \equiv \begin{array}[t]{lll}
  & (\etrans{\code{E}}(\Gamma_l) \land \ftrans{\mathcal{P}}(\Gamma_l,\Gamma_r)) & \lor \:
    (\etrans{E'}(\Gamma_r) \land \ftrans{\mathcal{P'}}(\Gamma_l,\Gamma_r)) \; \lor \\
  &     (\lnot \etrans{E}(\Gamma_l) \land \lnot \etrans{E'}(\Gamma_r)) & \lor \:
(\etrans{E}(\Gamma_l) \land \etrans{E'}(\Gamma_r))    
                                          \end{array}
% & \\
%       \mbox{where } \mathcal{A} & \equiv &
%             \begin{array}{ll}
%             &  (\etrans{E}(\Gamma_l) \land \ftrans{\mathcal{P}}(\Gamma_l,\Gamma_r)) \\
%       \lor  & (\etrans{E'}(\Gamma_r) \land \ftrans{\mathcal{P'}}(\Gamma_l,\Gamma_r)) \\
%       \lor  & (\lnot \etrans{E}(\Gamma_l) \land \lnot \etrans{E'}(\Gamma_r)) \\
%        \lor & (\etrans{E}(\Gamma_l) \land \etrans{E'}(\Gamma_r))    \end{array}
  \end{array} \]
  \caption{Translation of biprograms, excerpts}
  \label{fig:bi-trans}
\end{figure}

Translation of biprograms is described in Fig.~\ref{fig:bi-trans}.  The
translation function $\mathcal{B}$ takes a biprogram and a pair of contexts
$(\Gamma_l,\Gamma_r)$ to a WhyML program.  In addition to mapping \WhyRel\
identifiers to WhyML identifiers, contexts store information about the state
parameters on which the generated WhyML program acts.  Similar to
$\mathcal{B}$, the function $\mathcal{U}$ translates unary programs to WhyML
programs, $\mathcal{E}$, expressions to WhyML expressions, and $\mathcal{F}$,
a restricted set of relation formulas to WhyML expressions.  Biprograms don't
require the underlying unary programs to act on a disjoint set of variables;
however, this means that \WhyRel\ has to perform appropriate renaming during
translation.  Renaming is manifest in the translation of variable blocks
(\code{var x:T|x:T' in CC}), where the context $\Gamma_l$ (and
resp. $\Gamma_r$) is extended, $[\Gamma_l \mid \code{x} :\, x_l]$, mapping
\code{x} to a renamed copy $x_l$ (and resp. $\Gamma_r$ is extended with the
binding $\code{x} :\, x_r$).

In translating $\splitbi{C}{C'}$, the unary translations of $C$ and $C'$ are
sequentially composed.  Syncs $\syncbi{C}$ are handled similarly, as syntactic
sugar for $\splitbi{C}{C}$, except for the case of method calls.
Procedure-modular reasoning about relational properties is enabled by aligning
method calls which indicates that the relational spec associated with the
method is to be exploited.  \WhyRel\ will translate these to calls to the
appropriate WhyML product program, using a global method context ($\Phi$ in
Fig.~\ref{fig:bi-trans}).  Since translated product programs act on pairs of
states, the generated WhyML call takes $\Gamma_l.\code{st}$ and
$\Gamma_r.\code{st}$, names for left and right state parameters, as additional
arguments.

Product constructions for control flow statements require generating
additional proof obligations.  For aligned conditionals, \WhyRel\ introduces
an assertion that the guards are in agreement.  Lockstep aligned loops are
dealt with similarly; guard agreement must be invariant.  For conditionally
aligned loops, the generated loop body captures the pattern indicated by the
alignment guards $\mathcal{P} \!\mid\! \mathcal{P}'$: if the left
(resp. right) guard is true and $\mathcal{P}$ (resp. $\mathcal{P}'$) holds,
perform a left-only (resp. right-only) iteration; otherwise, perform a
lockstep iteration.  Adequacy is ensured by requiring the condition
$\mathcal{A}$ to be invariant.  This condition states that until
both sides terminate, the loop can perform a lockstep or a one-sided
iteration.  In relational region logic, the alignment guards $\mathcal{P}$ and
$\mathcal{P}'$ can be any relational formula.  However, the encoding of
conditionally aligned loops is in terms of a conditional that branches on
these alignment guards.  In Why3, this only works if $\mathcal{P}$ and
$\mathcal{P'}$ are restricted; for example, to not contain quantifiers.
\WhyRel\ supports alignment guards that include agreement formulas, one-sided
points-to assertions, one-sided boolean expressions, and the usual boolean
connectives.

\paragraph{Proof obligations for encapsulation.}  To ensure sound
encapsulation, \WhyRel\ performs an analysis on source programs.  This
analysis includes two parts: a static check to ensure client programs don't
directly write to variables in a module's boundary; and the generation of
intermediate assertions that express disjointness between the footprints of
client heap updates and regions demarcated by module boundaries.  For modules
with public/private invariants, \WhyRel\ additionally generates a lemma which
states that the module's boundary frames the invariant, i.e., the invariant
only depends on locations expressed by the boundary.  The same is done with
coupling relations, for which we need to consider boundaries of both modules
being related.  A technical condition of relational region logic requiring
boundaries grow monotonically as computation proceeds is also ensured by
introducing appropriate postconditions in generated programs.

\section{Evaluation}
\label{sec:evaluation}

We evaluate \WhyRel\ via a series a case studies, representative of the
challenge problems highlighted at the outset of this article.  Examples
include representation independence, optimizations such as loop
tiling~\cite{BartheCK13}, and others from recent literature on relational
verification (including~\cite{EilersMH18} and~\cite{NaumannISOLA20}).  Some,
like those described in Sec.~\ref{sec:alignment}, deal with reasoning in terms
of varying alignments including data-dependent ones. Our representation
independence examples include showing equivalence of Dijkstra's single-source
shortest-paths algorithm linked against two implementations of priority
queues, which requires reasoning about fine-grained couplings between pointer
structures; and Kruskal's minimum spanning tree algorithm linked against
different modules implementing union-find, which requires couplings equating
the partitions represented by the two versions.  For all examples, VCs are
discharged using the SMT solvers Alt-Ergo, CVC4, and Z3.  Replaying proofs of
most developments using Why3's saved sessions feature takes less than 30
minutes on a machine with an Intel Core i5-6500 processor and 32 gigabytes of
RAM.
% Most developments take less than 30 minutes to fully verify in Why3 on a
% machine with an Intel Core i5-6500 processer and 32 gigabytes of RAM.
% CHECK RN: addressed
% \dn{Say the artifact includes saved sessions and these timings are for replays?}

A primary goal of this work is to investigate whether verifying relational
properties of heap manipulating programs can be performed in a manner
tractable to SMT-based automation, and for the most part, we believe \WhyRel\
provides a promising answer.  The tool serves as an implementation of
relational region logic and demonstrates that even its additional proof
obligations for encapsulation can be encoded using first-order assertions.  In
fact, exploration of case studies using \WhyRel\ was instrumental in designing
proof rules of relational region logic.
% FUTURE In final version we can say: in fact, exploration of case studies using WhyRel
% was instrumental in designing the relational proof rules
% (cf. early ArXiv version vs final).

Reasoning about heap effects \`a la region logic is generally simple and VCs
get discharged quickly using SMT.  However, technical lemmas \WhyRel\
generates which pertain to showing that module boundaries frame private
invariants and couplings require considerable manual effort to prove.  These
lemmas usually involve reasoning about image expressions, which involve
existentials and nontrivial set operations on regions.  Given our encoding of
states and regions, SMT solvers seem to have difficulties solving these goals.
Manual effort involves applying a series of Why3 transformations (or proof
tactics) and introducing intermediate assertions.  We conjecture that the
issue can be mitigated by using specialized solvers~\cite{RosenbergBN12} or
different heap encodings~\cite{Schmid21}.

Another issue with our encoding of typed program states is the generation of a
large number of VCs related to well-formedness of states.  These account for a
substantial fraction of proof replay time.  Why3 programs act directly on our
minimally-typed state representation and each heap update needs to preserve an
invariant that specifies constraints on the types of allocated references (see
Fig.~\ref{fig:state-encoding}).  Using Why3's support for module
abstraction~\cite{Why3Abstraction} may ameliorate this issue.  An alternative
is to use assumptions, which can be justified by correctness of the \WhyRel\
type checker and translator.\footnote{The Boogie verification language
  provides ``free requires'' and ``free ensures'' syntax for just such
  assumptions.}

Apart from these challenges related to verification, we note that specs in
region logic tend to be verbose when compared to other formalisms such as
separation logic~\cite{RegLogJrnI}.

\section{Related work}
\label{sec:related-work}

\WhyRel\ is closely modeled on relational region logic, developed
in~\cite{BNNN22}.  That paper provides a high-level overview of \WhyRel{},
using a small set of examples verified in the tool to motivate aspects of the
formal logic; but it doesn't give a full presentation of the tool or go into
details about the encoding.  The paper provides comprehensive soundness proofs
of the logic and shows how the VCs \WhyRel\ generates and the checks it
performs correspond closely to obligations of relational proof rules.  The
paper builds on a line of work on region
logic~\cite{RegLogJrnI,RegLogJrnII,BanerjeeNN18}.  The VERL tool implements an
early version of unary region logic without encapsulation and was used to
evaluate a decision procedure for regions~\cite{RosenbergBN12}.

For local reasoning about pointer programs, separation logic is an effective
and elegant formalism.  For relational verification, ReLoC~\cite{FruminKB18},
based on the Iris separation logic and built in the Coq proof assistant
supports, apart from many others, language features such as dynamic allocation
and concurrency.  However, we are unaware of auto-active relational verifiers
based on separation logic.
% that feature encapsulation or take an auto-active view on verification.  [Iris does encap]

Alignments for relational verification have been explored in various contexts.
In \WhyRel{}, the biprogram syntax captures alignment based on control flow, but
also caters to data-dependent alignment of loops through the use of alignment
guards (as discussed in Sec.~\ref{sec:alignment}).  Churchill et
al.~\cite{ChurchillP0A19} develop a technique for equivalence checking by
using data dependent alignments represented by control flow automata which
they use to prove correctness of a benchmark of vectorizing compiler
transformations and hand-optimized code.  Unno et
al.~\cite{UnnoTerauchiKoskinen21} address a wide range of relational problems
including $k$-safety and co-termination, expressing alignments and invariants
as constraint satisfaction problems they solve using a CEGIS-like technique.
Their work is applied to benchmarks proposed by Shemer et
al.~\cite{ShemerGSV19} who develop a technique for equivalence and regression
verification.  Both the above works represent alignments as transition systems
and perform inference of relational invariants and alignment conditions.
Inference relies on solvers and therefore programs need to be restricted so
they are amenable to these solvers.  A promising approach by Barthe et
al.~\cite{BartheTraceLogic19} reduces relational verification to proving
formulas in trace logic, a multi-sorted first-order logic using first-order
provers.  In trace logic, conditions can be expressed on traces including
relationships between different time points without recourse to alignment per
se.

Sousa and Dillig develop Descartes~\cite{SousaD16} for reasoning about
$k$-safety properties of Java programs automatically using implicit product
constructions and in a logic they term Cartesian Hoare logic.  Their work is
furthered by Pick et al.~\cite{PickFG18} who develop novel techniques for
detecting alignments.  The REFINITY~\cite{Steinhofel20} workbench based on the
interactive KeY tool can be used to reason about transformations of Java
programs; heap reasoning relies on dynamic frames and relational verification
proceeds by considering \emph{abstract programs}.  Other related tools include
SymDiff~\cite{LahiriHKR12} which is based on Boogie and can modularly reason
about program differences in a language-agnostic way, and
LLR\^eve~\cite{KieferKU18} for regression verification of C programs.  Eilers
et al.~\cite{EilersMH20} develop an encoding of product programs for
noninterference that facilitates procedure-modular reasoning.  They verify a
large collection of benchmark examples using the VIPER toolchain.

\section{Conclusion}
\label{sec:conclusion}

In this paper we present \WhyRel{}, a prototype for relational verification of
pointer programs that supports dynamic framing and state-based encapsulation.
The tool faithfully implements relational region logic and demonstrates how
its proof obligations, including those related to encapsulation, can be
encoded in a first-order setting.  We've performed a number of representative
examples in \WhyRel\, leveraging support Why3 provides for SMT, and believe
these demonstrate the amenability of region logic, and its relational variant,
to automation.

\subsubsection{Acknowledgments}

We thank the anonymous TACAS reviewers and artifact evaluators for their
thorough feedback and suggestions which have led to major improvements in this
paper.  We thank Seyed Mohammad Nikouei who built an initial version of
\WhyRel\ which helped guide the design of the current version.  Nagasamudram
and Naumann were partially supported by NSF award 1718713. Banerjee's research
was based on work supported by the NSF, while working at the Foundation. Any
opinions, findings, and conclusions or recommendations expressed in this
article are those of the authors and do not necessarily reflect the views of
the NSF.

\subsubsection{Data Availability Statement}

Sources for \WhyRel\ and all examples performed using the tool are
available in Zenodo with the identifier
\url{https://doi.org/10.5281/zenodo.7308342}~\cite{WhyRelArtifact}.

%\bibliographystyle{splncs04}
%\bibliography{../relRLabs/paper}

\end{document}